\documentclass[prl,twocolumn,floatix,amssymb,amsmath,superscriptaddress,letterpaper]{revtex4}
\usepackage{graphicx}
\usepackage{epsfig}
\usepackage{amsmath}
\usepackage{amssymb}
\usepackage{amsfonts}
\usepackage{mathrsfs}
\usepackage{theorem}
\usepackage{bm}
\usepackage{url}
\usepackage[T1]{fontenc}
\usepackage{csquotes}
\usepackage{natbib}
\usepackage{float}
\usepackage{appendix}
\usepackage{subfigure}
\MakeOuterQuote{"}

\usepackage{dcolumn}
\usepackage{color}
\usepackage{color}

\begin{document}

\title{Contextuality of Identical Particles}

\author{Pawe{\l} Kurzy\'{n}ski}
\email{pawel.kurzynski@amu.edu.pl}
\affiliation{Faculty of Physics, Adam Mickiewicz University, Umultowska 85, 61-614 Pozna\'{n}, Poland}
\affiliation{Centre for Quantum Technologies, National University of Singapore, 3 Science Drive 2, 117543 Singapore, Singapore}

%%%%%%%%%%%%%%%%%%%%%%%%%%%%%%%%%%%%%%%%%%%%%%%%%

\begin{abstract}
A generalisation of quantum contextuality to the case of many indentical particles is presented. The model consists of a finite collection of modes that can be occupied by N particles, either bosons or fermions. Measurement scenarios allow one to measure occupation of each mode in at least two different measurement contexts. The system is said to be non-contextual if the occupation numbers can be assigned to modes in each measurement scenario. The assignment is done under the non-contextuality assumption, i.e., an occupation number assigned to a mode does not depend on a scenario in which this mode is measured. In addition, the total number of particles has to be conserved, therefore the sum of occupation numbers in each measurement context is equal to $N$. For $N=1$ the model does not differ from the standard contextuality scenario. However, for $N>1$ the problem reveals new complex features. In particular, it is shown that a type of contextuality exhibited by the system (state-dependent, state-independent, or non-contextual) depends on the type and the number of particles. Further properties of this model and open problems are also discussed.  
\end{abstract}

%%%%%%%%%%%%%%%%%%%%%%%%%%%%%%%%%%%%%%%%%%%%%%%%%

\maketitle

%%%%%%%%%%%%%%%%%%%%%%%%%%%%%%%%%%%%%%%%%%%%%%%%%

{\it Introduction.} Typical tests of hidden-variable (HV) models are derived either for a single indivisible system \cite{KS} or assume that subsystems are distinguishable and can be addressed individually \cite{Bell}. On the other hand, majority of common physical systems consist of many indistinguishable parts. Therefore, derivation of HV tests for collections of indistinguishable objects is necessary to understand nonclassical phenomena in realistic scenarios. 

Although the HV problem has been extensively studied in various physical systems, the case of identical particles occupying a family of local modes is still barely explored \cite{AdanMarcelo,Us1,Com1,Us2,Liang}. The goal of this work is to provide a method to test a HV description of measurements done on a collection of bosons or fermions which occupy some number of modes. In particular, the aim is to generalise the Kochen-Specker theorem \cite{KS} and the concept of quantum contextuality to scenarios in which HV correspond to an assignment of particle occupation numbers. 

The standard way of proving the Kochen-Specker (KS) theorem \cite{KS}, which due to historical reasons is also known as the Bell-Kochen-Specker theorem \cite{BKS}, is based on an assignment of logical values to a finite set of projectors $\{{\mathcal P_i}\}_{i \in K}$ in some finite-dimensional Hilbert space. For each subset of mutually orthogonal projectors $\{{\mathcal P_j}\}_{j\in S_{\mathcal A} \subset K}$ having the resolution of identity one assigns the value $v\left({\mathcal P_k}\right)=1$ to exactly one projector and the value $v\left({\mathcal P_{l\neq k}}\right)=0$ to all the remaining projectors from this subset. In this way one constructs a classical-like description of a quantum measurement ${\mathcal A}$ in which an observed event corresponds to the projector that was assigned 1. In addition, the non-contextuality assumption states that the value assigned to a projector is the same in every subset. One also requires that each projector belongs to more than one mutually orthogonal subset. The goal is to arrive at a contradiction, i.e., to show that in some subset all projectors will be assigned 0, or that more than one projector is assigned 1. In this case the system is said to be contextual. 

The above contradiction occurs for any state and therefore it is an example of the state-independend contextaulity (SIC). However, there are two ways one can arrive at SIC. The first one, discussed above, relies on finding a set of projectors that cannot be assigned values 0 and 1 in a way not leading to a contradiction. Such sets are known in the literature as the KS-sets \cite{KSset}. The other approach relies on an inequality for a set of projectors for which 0-1 assignments may exist. These assignments are used to derive a non-contextuality bound. However, due to the properties of projectors in the set, the bound is violated by any quantum state. Such sets of observables are known as the SIC-sets \cite{SIC1,SIC2,SIC3}. 

There is also another type of contextuality that occurs for specific states (state-dependent contextuality). In this case possible assignments provide a non-contextual model for some subset of states. The goal is to look for a set of projectors minimizing the subset of non-contextual states and for states lying outside of this subset \cite{KCBS,CSW}. 

%%%%%%%%%%%%%%%%%%%%%%%%%%%%%%%%%%%%%%%%%%%%%%%%%

{\it The general idea.} The KS scenario can be generalized in the following way. Instead of a set of events, one considers a set of modes that are populated by $N$ noninteracting particles. Let me first discuss a single-particle case ($N=1$) and a system consisting of $d$ orthogonal modes. One starts with a finite set of modes  $\{{\mathcal M_i}\}_{i \in K}$ from which one chooses a subset $\{{\mathcal M_j}\}_{j\in S_{\mathcal A} \subset K}$ consisting of $d$ mutually orthogonal ones. A measurement ${\mathcal A}$, corresponding to this subset, reveals that a particle can be found in exactly one mode, hence one can consider a classical-like description in which one assigns $v\left({\mathcal M_k}\right)=1$ to the occupied mode and  $v\left({\mathcal M_{l\neq k}}\right)=0$ to unoccupied modes. Moreover, if one also assumes non-contextuality, then the value assigned to a mode is the same in every subset in which this mode occurs. The goal is to show that such an assignment is not possible. 

The single-particle example is basically the same as the standard KS scenario. Perhaps, the only difference is in the interpretation of the contradiction at which one arrives. In the original case one has a logical contradiction that two exclusive events happen at the same moment, or that no event happens. In the particle-mode example the contradiction has a more physical basis and corresponds to a lack of the particle number conservation.

The situation becomes more interesting when $N>1$ and particles are indistinguishable. The case of many distinguishable particles can be reduced to many single-particle cases, because mode occupation can be considered for each type of particle separately, hence it does not differ from the standard scenario. However, the problem reveals new complex features when the particles are bosons or fermions. In case of bosons the values assigned to modes are $v_B\left({\mathcal M_i}\right)=0,1,2,\dots$, whereas for fermions they are $v_F\left({\mathcal M_i}\right)=0,1$. In addition, for each subset of mutually orthogonal modes having the resolution of identity the following holds
\begin{equation}\label{conservation}
\sum_{i\in S_{\mathcal A}} v_B\left({\mathcal M_i}\right) = \sum_{i\in S_{\mathcal A}} v_F\left({\mathcal M_i}\right) = N.
\end{equation}

Just like in the standard KS scenario, the problem can be formulated for a system having $d \geq 3$ distinguishable modes. For bosons $N$ is arbitrary, but for fermions $N\leq d$.

It is also important to mention the dimensionality of the Hilbert space considered in the problem. The number of elements in each subset to which one assigns values is equal to the number $d$ of orthogonal modes having the resolution of identity. It does not depend on the number $N$ of particles. This is different from the dimensionality of the corresponding Fock space, which equals $\frac{d!}{N!(N-d)!}$ for fermions and $\frac{(d+N-1)!}{N!(d-1)!}$ for bosons, for which the standard KS scenario can be considered \cite{AdanMarcelo}.

\begin{figure}[t]
\includegraphics[width=0.5 \textwidth,trim=4 4 4 4,clip]{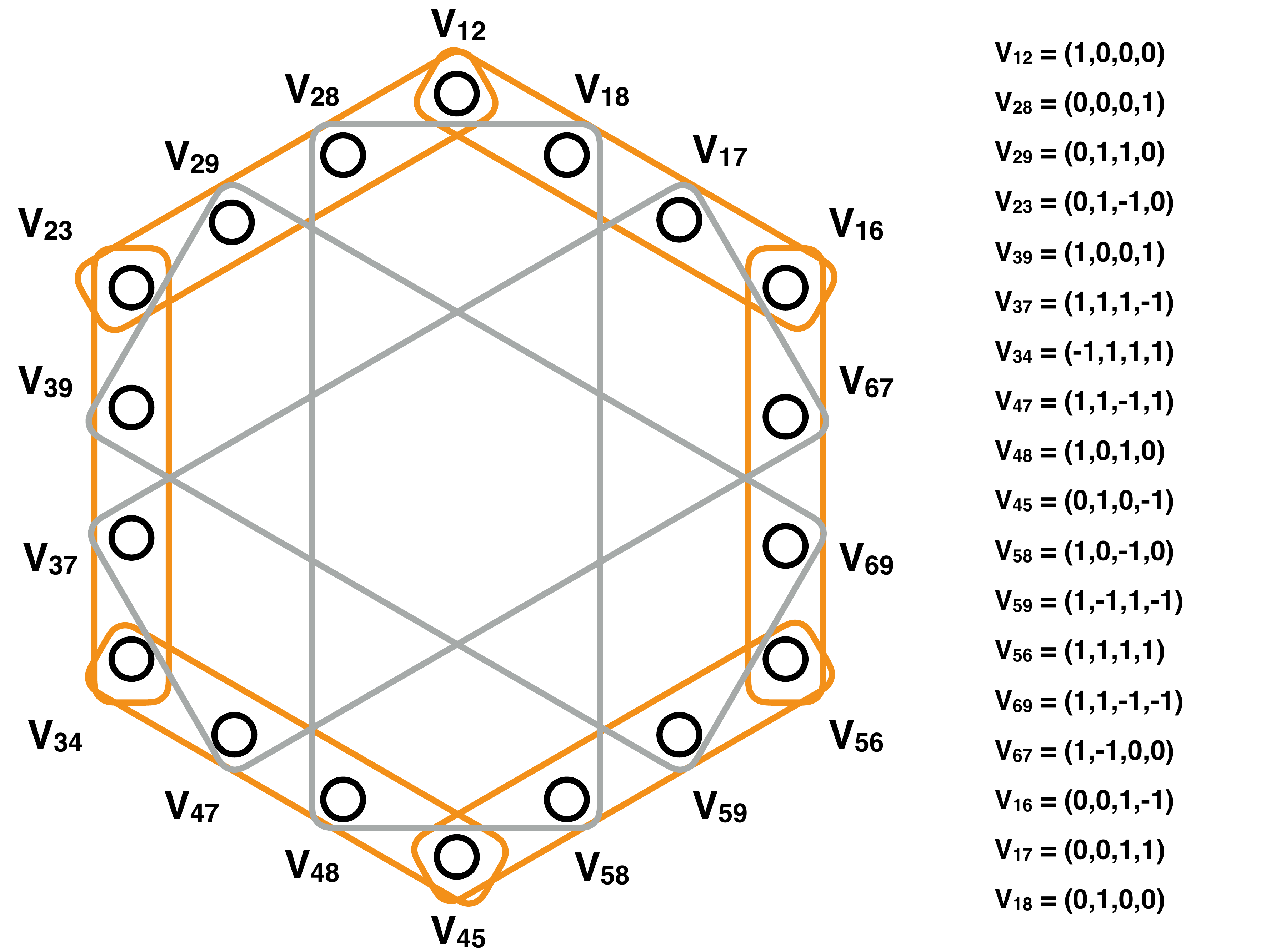}
\caption{ A hypergraph from Ref. \cite{Adan2008} representing measurements in the 18-projector KS scenario in dimension $d=4$ \cite{Adan18}. Hyperedges (grey and orange rectangular areas) correspond to measurement contexts, i.e., sets of mutually orthogonal projectors (modes). The projectors in the right column are unnormalized.}
\label{fig1}
\end{figure}

In the rest of this work I focus on the 18-projector KS-set \cite{Adan18,Adan2008} and on its properties in $N>1$ scenarios, however the model developed here allows one to consider any set of modes. The reason to study the 18-projector set is to show how the familiar model changes when more particles are considered. For this set $d=4$ (see Fig. \ref{fig1}). Following the notation in \cite{Adan2008}, the modes are labeled $v_{ij}$, where $i$ and $j$ denote the measurement contexts to which the mode belongs to. For $N=1$ one arrives at a contradiction. This is because there are 9 measurement contexts and since any proper value assignment requires that in each context exactly one mode is assigned 1, the sum of assignments over all contexts must be equal to 9. However, in this sum each mode appears twice, which implies that the sum must give an even number -- one arrives at a contradiction. In the following sections I discuss the case $N>1$ for fermions and bosons. 

%%%%%%%%%%%%%%%%%%%%%%%%%%%%%%%%%%%%%%%%%%%%%%%%%

%%%%%%%%%%%%%%%%%%%%%%%%%%%%%%%%%%%%%%%%%%%%%%%%%

%%%%%%%%%%%%%%%%%%%%%%%%%%%%%%%%%%%%%%%%%%%%%%%%%

{\it Fermions.} For $d=4$ the total number of fermions can be $N=0,1,\ldots,4$. The case $N=0$ is obviously non-contextual since one assigns 0 to each mode. As discussed above, the case $N=1$ corresponds to a standard KS scenario and the 18 modes form a KS-set. The case $N=2$ is the first non-trivial extension of the KS scenario to more than one particle and is going to be discussed in a moment. The case $N=3$ has an interesting interpretation because it is the standard KS scenario in which 0's are swapped with 1's. Physically, this scenario can be interpreted as a search for a non-contextual model for a single hole. The symmetry between particles and holes implies that for $N=3$ the 18 modes form a KS-set, because they form a KS-set for $N=1$. In general, in fermionic case one can speak of N particles or alternatively of $d-N$ holes. Therefore, $N=4$ is equivalent to $N=0$ and the system is non-contextual. Because of the above, the generalisation of contextuality to many fermions is non-trivial for $d > 3$, which is an additional reason why the 18-projector example in $d=4$ dimensions is discussed in this work.

%%%%%%%%%%%%%%%%%%%%%%%%%%%%%%%%%%%%%%%%%%%%%%%%%

\begin{figure}[t]
\includegraphics[width=0.5 \textwidth]{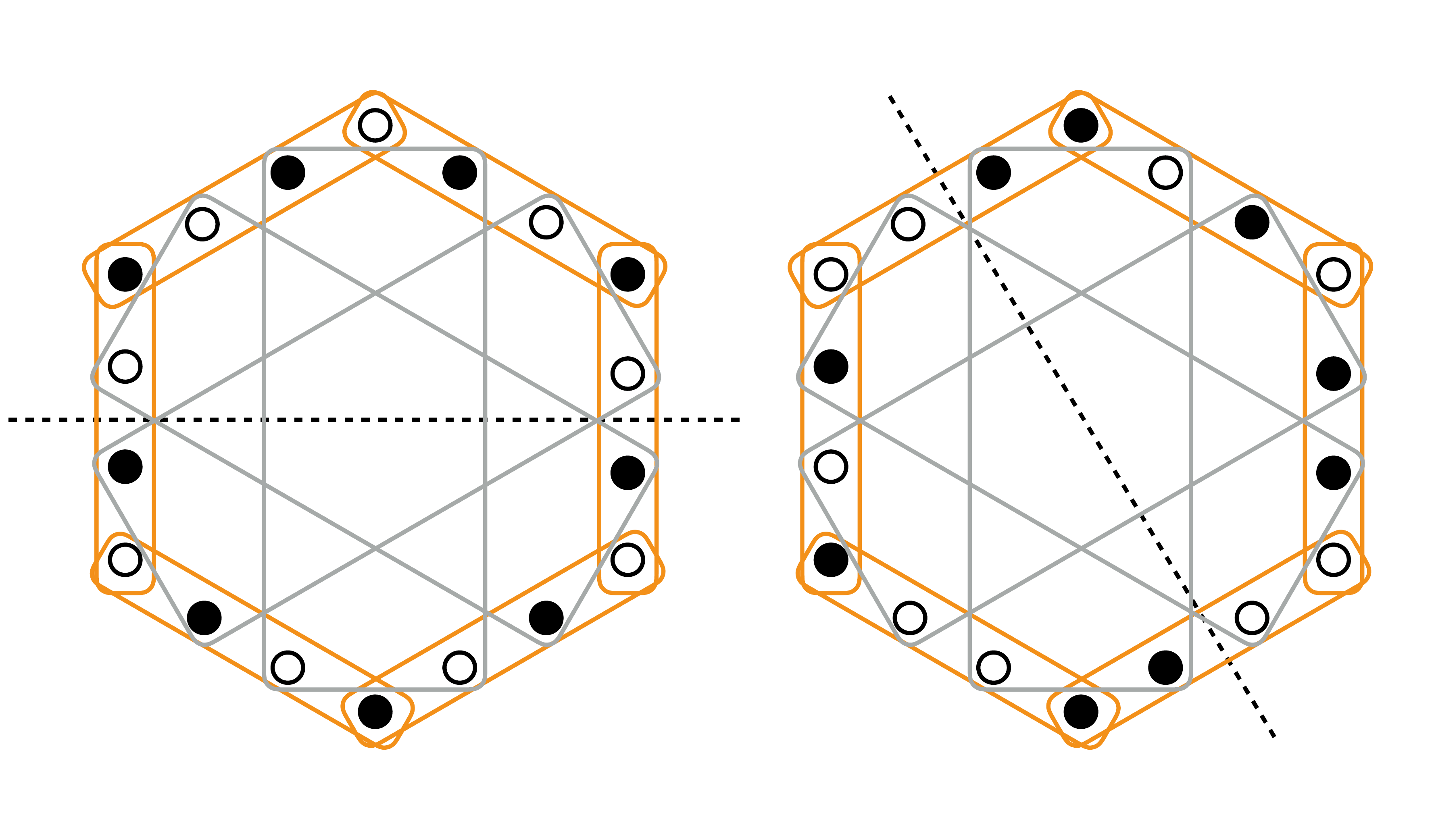}
\caption{ Possible mode-assignments of two fermions in the 18-projector KS set. Filled nodes correspond to particles and empty nodes correspond to holes. These assignments exhibit a symmetry between particles and holes, i.e., if one exchanges particles with holes one obtains the same assignment (up to the reflection denoted by the dashed lines).}
\label{fig2}
\end{figure}

%%%%%%%%%%%%%%%%%%%%%%%%%%%%%%%%%%%%%%%%%%%%%%%%%

For $N=2$ fermions one can find an assignment to the 18-projector set fulfilling the criteria of non-contextuality and particle number conservation (Fig. \ref{fig2}). Therefore, for $N=2$ the 18 modes do not form a KS-set. They do not form a simple SIC-set either. The SIC-set constitutes an operator $C_{SIC}$ which in its simplest form is a sum of projectors onto all modes in the set. The corresponding SIC inequality bounds the expectation value of the SIC-operator by the sum of occupation numbers over all modes in a non-contextual model. Because of the symmetry between particles and holes, any non-contextual model assigns exactly 9 particles and 9 holes to all 18 modes (see Fig. \ref{fig2}). Therefore, the SIC-operator is bounded by 9. 

In order to evaluate the average value of the expectation of the SIC-operator note that 18 modes constitute 9 contexts, each having a resolution of identity, and each mode appears in exactly two contexts. Therefore, $C_{SIC}=\frac{9}{2}\hat N$, where $\hat N$ is an operator of the total number of particles. As a result, for any state of two fermions (or bosons) $|\psi\rangle$ one has $\langle \psi |C_{SIC}|\psi\rangle=9$ and the corresponding SIC inequality cannot be violated. However, the possibility of violation of some other SIC inequality based on an operator which is not a simple sum of projectors onto all modes was not excluded here and remains an open problem.

The assignments discussed above exhibit a symmetry between particles and holes, which suggests that in order to observe contextuality one has to find a state which brakes this symmetry. Therefore, one has to consider a state-dependent version of contextuality. Indeed, for $N=2$ one can find a Hardy-like proof \cite{Hardy} of quantum contextuality \cite{HardyC1,HardyC2}.

%%%%%%%%%%%%%%%%%%%%%%%%%%%%%%%%%%%%%%%%%%%%%%%%%

 {\it Proof of Hardy-like contextuality for two fermions.} Let $f^{\dagger}_{ij}$ be an operator creating a fermion in the mode $v_{ij}$ (see Fig. \ref{fig1}). These operators obey standard fermionic anticommutation rules $f_{ij}^{\dagger}f_{kl}^{\dagger}+f_{kl}^{\dagger}f_{ij}^{\dagger}=0$. Next, consider the state
\begin{equation}\label{h6}
|\psi\rangle = f^{\dagger}_{67}f^{\dagger}_{69}|0\rangle.
\end{equation} 
This state has a well defined fermionic occupation in the measurement context 6. Then, consider the same state in the measurement contexts 3, 7 and 9
\begin{eqnarray}
|\psi\rangle &=& \left( \frac{f^{\dagger}_{39}f^{\dagger}_{23}}{2\sqrt{2}}   + \frac{f^{\dagger}_{37}f^{\dagger}_{23}}{4} - \frac{f^{\dagger}_{37}f^{\dagger}_{39}}{4}  - \frac{3f^{\dagger}_{34}f^{\dagger}_{23}}{4} \right. \nonumber \\ &+& \left. \frac{f^{\dagger}_{34}f^{\dagger}_{39}}{4} - \frac{f^{\dagger}_{34}f^{\dagger}_{37}}{2\sqrt{2}} \right)|0\rangle, \label{h3} \\
|\psi\rangle &=& \left( \frac{f^{\dagger}_{67}f^{\dagger}_{37}}{2}  + \frac{f^{\dagger}_{67}f^{\dagger}_{47}}{2}  + \frac{f^{\dagger}_{17}f^{\dagger}_{67}}{\sqrt{2}}  \right)|0\rangle, \label{h7} \\
|\psi\rangle &=& \left( \frac{f^{\dagger}_{69}f^{\dagger}_{29}}{2}  - \frac{f^{\dagger}_{69}f^{\dagger}_{39}}{2}  - \frac{f^{\dagger}_{69}f^{\dagger}_{59}}{\sqrt{2}}  \right)|0\rangle. \label{h9}
\end{eqnarray}

One begins by making a measurement in the context 3. According to Eq. (\ref{h3}) there is a probability $\frac{1}{16}$ that one finds particles in modes $v_{37}$ and $v_{39}$. If this happens, context 3 has the following assignment 
\begin{equation}
C_3=\{v_{34}=0,v_{37}=1,v_{39}=1,v_{23}=0\}. 
\end{equation}

Next, from Eqs. (\ref{h7}), (\ref{h9}) and (\ref{h6}) one has 
\begin{eqnarray}
C_7 &=& \{v_{17}=0,v_{67}=1,v_{47}=0,v_{37}=1\}, \\
C_9 &=& \{v_{69}=1,v_{59}=0,v_{39}=1,v_{29}=0\}, \\
C_6 &=& \{v_{16}=0,v_{67}=1,v_{69}=1,v_{56}=0\}. 
\end{eqnarray} 
Note, that $C_7$ and $C_9$ are implied by $C_3$ and Eqs. (\ref{h7}) and (\ref{h9}), i.e., there is only a single term in (\ref{h7}) and (\ref{h9}) that contains particle in $v_{37}$ and $v_{39}$, respectively. The assignment $C_6$ is implied by $C_7$ and $C_9$. In addition, it is confirmed by the state $(\ref{h6})$.

Because $v_{17}$ and $v_{16}$ has been already assigned 0, one gets
\begin{equation}
C_1 = \{v_{12}=1,v_{18}=1,v_{17}=0,v_{16}=0\}.
\end{equation}
Similarly, $v_{56}=0$ and $v_{59}=0$ imply
\begin{equation}
C_5 = \{v_{56}=0,v_{59}=0,v_{58}=1,v_{45}=1\}.
\end{equation}
One can follow the chain of implications to obtain
\begin{equation}
C_8 = \{v_{18}=1,v_{58}=1,v_{48}=0,v_{28}=0\}.
\end{equation}

%%%%%%%%%%%%%%%%%%%%%%%%%%%%%%%%%%%%%%%%%%%%%%%%%

\begin{figure}[t]
\includegraphics[width=0.5 \textwidth]{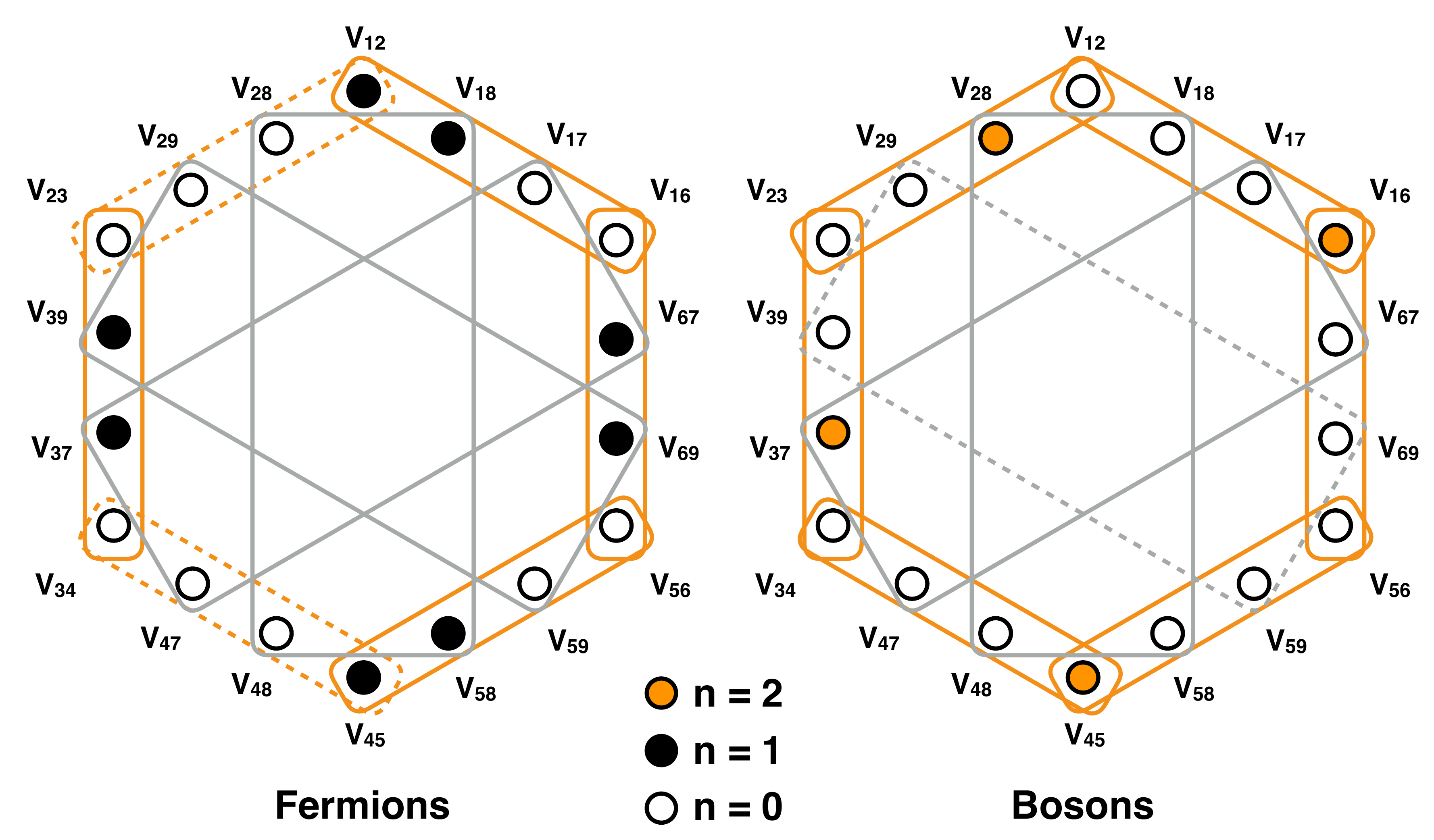}
\caption{Mode assignments and contradictions in the Hardy-like proofs of contextuality for two identical particles. The contexts with contradictory assignments are denoted by dashed hyperedges.}
\label{fig3}
\end{figure}

%%%%%%%%%%%%%%%%%%%%%%%%%%%%%%%%%%%%%%%%%%%%%%%%%

At this point all the modes have been already assigned an occupation number. However, one obtains a contradiction (see Fig. \ref{fig3}) because there are two contexts that are assigned only a single particle
\begin{eqnarray}
C_2 &=& \{v_{23}=0,v_{29}=0,v_{28}=0,v_{12}=1\}, \\
C_4 &=& \{v_{45}=1,v_{48}=0,v_{47}=0,v_{34}=0\}.
\end{eqnarray} 

Finally, note that the above contradiction can be obtained for any state of the form $f^{\dagger}_{m}f^{\dagger}_{n}|0\rangle$, where the modes $m$ and $n$ are orthogonal. One simply needs to find an arbitrary unitary operation which transforms $f^{\dagger}_{67} \rightarrow f^{\dagger}_{m}$ and $f^{\dagger}_{69} \rightarrow f^{\dagger}_{n}$ and apply it to all 18 modes. Next, one needs to follow similar steps as above.

%%%%%%%%%%%%%%%%%%%%%%%%%%%%%%%%%%%%%%%%%%%%%%%%%

{\it Bosons.} The bosonic case for $N=0,1,2$ resembles the fermionic one. The 18-mode set for $N=0$ bosons is trivially non-contextual and $N=1$ corresponds to the standard KS scenario. The two-fermion occupation number assignments from Fig. \ref{fig2} can also describe a system of bosons and therefore the $N=2$ bosonic case does not provide a typical KS contradiction and the 18 modes do not form a KS-set. Due to the reasons discussed in the previous sections, the 18 modes do not form a simple SIC-set either. However, one can find a Hardy-like proof by exploiting the bosonic properties of the two particles.

%%%%%%%%%%%%%%%%%%%%%%%%%%%%%%%%%%%%%%%%%%%%%%%%%

 {\it Proof of Hardy-like contextuality for two bosons.} Let $b^{\dagger}_{ij}$ be an operator creating a boson in the mode $v_{ij}$. The bosonic creation operators obey standard commutation rules $b_{ij}^{\dagger}b_{kl}^{\dagger}-b_{kl}^{\dagger}b_{ij}^{\dagger}=0$. Next, consider the state
\begin{equation}\label{h16}
|\psi\rangle = \frac{b^{\dagger^2}_{16}}{\sqrt{2}}|0\rangle,
\end{equation} 
which has a well defined bosonic occupation in contexts 1 and 6. The operator $b^{\dagger}_{16}$ has the following representation in the basis of operators from the context 4
\begin{equation}\label{h16c4}
b^{\dagger}_{16}=\frac{b^{\dagger}_{45}}{2}+\frac{b^{\dagger}_{48}}{2}-\frac{b^{\dagger}_{47}}{\sqrt{2}}.
\end{equation}
This implies that the state $|\psi\rangle$ has an equivalent form
\begin{eqnarray}
|\psi\rangle &=& \left(\frac{b^{\dagger^2}_{45}}{4\sqrt{2}}+\frac{b^{\dagger^2}_{48}}{4\sqrt{2}}-\frac{b^{\dagger^2}_{47}}{2\sqrt{2}} + \frac{b^{\dagger}_{45}b^{\dagger}_{48}}{4\sqrt{2}} \right. \nonumber \\ 
&-& \left. \frac{b^{\dagger}_{45}b^{\dagger}_{47}}{4}-\frac{b^{\dagger}_{47}b^{\dagger}_{48}}{4}\right)|0\rangle \label{psib4}
\end{eqnarray}
and that there is a probability $\frac{1}{16}$ to measure two particles in mode $v_{45}$. If this happens one has the following assgnments
\begin{eqnarray}
C_4 &=& \{v_{45}=2,v_{48}=0,v_{47}=0,v_{34}=0\}, \\
C_5 &=& \{v_{56}=0,v_{59}=0,v_{58}=0,v_{45}=2\}.
\end{eqnarray}
In addition, the initial assumption about the state (\ref{h16}) implies
\begin{eqnarray}
C_1 &=& \{v_{12}=0,v_{18}=0,v_{17}=0,v_{16}=2\}, \\
C_6 &=& \{v_{16}=2,v_{67}=0,v_{69}=0,v_{56}=0\}. 
\end{eqnarray}
From the above four assignments one finds that since $v_{17}=v_{67}=v_{47}=0$ and $v_{18}=v_{58}=v_{48}=0$  then
\begin{eqnarray}
C_7 &=& \{v_{17}=0,v_{67}=0,v_{47}=0,v_{37}=2\}, \\
C_8 &=& \{v_{18}=0,v_{58}=0,v_{48}=0,v_{28}=2\}.
\end{eqnarray}
Next, since $v_{28}=v_{37}=2$ one gets
\begin{eqnarray}
C_2 &=& \{v_{23}=0,v_{29}=0,v_{28}=2,v_{12}=0\}, \\
C_3 &=& \{v_{34}=0,v_{37}=2,v_{39}=0,v_{23}=0\}. 
\end{eqnarray}

The occupation numbers are already assigned to all the modes, however one obtains a contradiction (see Fig. \ref{fig3}) since
\begin{equation}\label{bcontr}
C_9 = \{v_{69}=0,v_{59}=0,v_{39}=0,v_{29}=0\}.
\end{equation}
Just like in the case of two fermions, the proof can be adopted to any state of the form $\frac{b_j^{\dagger^2}}{\sqrt{2}}|0\rangle$.

%%%%%%%%%%%%%%%%%%%%%%%%%%%%%%%%%%%%%%%%%%%%%%%%%

 {\it Contextuality of $N>2$ bosons.} The above proof of Hardy-like contextuality can be easily generalised to the case of an arbitrary $N>0$ number of bosons. One starts with a state  
\begin{equation}\label{h16N}
|\psi\rangle = \frac{b^{\dagger^N}_{16}}{\sqrt{N!}}|0\rangle,
\end{equation} 
and measures the occupation numbers in the context 4. From the Eq. (\ref{h16c4}) one finds that the probability of finding $N$ particles in mode $v_{45}$ is $\frac{1}{4^N}$. If this happens, one follows the same steps as in the case of $N=2$. This time in all the assignments there is $N$ instead of 2. Finally, one arrives at (\ref{bcontr}).

The Hardy-like proof shows that there is the state-dependent contextuality in the 18-mode scenario for any $N>0$. However, the lack of the state-independent contextuality was only shown for $N=2$. It is possible that for some $N>2$ the bosonic occupation numbers of 18 modes cannot be assigned in a non-contextual way. This would require further research, since for large numbers of particles there are many ways in which the particles can be distributed between the modes. Note, that a non-contextual assignment of occupation numbers is an integer programming problem. The integer programming is in general NP-hard \cite{Papadimitriou} and it would be interesting to investigate which integer programming problems can be reduced to a non-contextual assignment of occupation numbers. 

%%%%%%%%%%%%%%%%%%%%%%%%%%%%%%%%%%%%%%%%%%%%%%%%%

{\it Conclusions.} In this work I presented a generalisation of quantum contextuality  to systems of identical particles in which one cannot assign particle occupation numbers to modes in a non-contextual way and under the particle number conservation assumption. In particular, for the system of 18 modes that can be grouped in 9 contexts of 4 mutually orthogonal ones (corresponding to a single-particle Hilbert space of dimension 4) I showed that the contextuality depends on the number and on the type of particles. The bosonic (fermionic) case $N=0$ ($N=0,4$) is trivially non-contextual. The case $N=1$ ($N=1,3$) exhibits state-independent contextuality. For the case $N=2$ there is no state-independent contextuality (for both bosons and fermions) in the form of a KS-set or a simple SIC-set, however a state-dependent Hardy-like proof of contextuality can be found. This proof also works for $N>2$ bosons. The possibility of state-independent contextuality in multi-bosonic systems remains to be investigated, however this task may not be simple since it seems to be related to the integer programming problem that is known to be hard.

%%%%%%%%%%%%%%%%%%%%%%%%%%%%%%%%%%%%%%%%%%%%%%%%%

{\it Major open problems.} The results reported in this work lead to a number of open questions. Apart from the ones mentioned above, the other major problems are the following: Can one find the state-independent contextuality of bosons for $N>1$ (fermions for $d-1>N>1$) and for an arbitrary $d\geq 3$ ($d \geq 4$)? For a given $d$ and $N$ what is the minimal number of modes that lead to the state-dependent and state-independent contextuality? The Hardy-like proof is probabilistic, but is it possible to find a Greenberger-Horne-Zeilinger-like (GHZ-like) \cite{GHZ} proof of contextuality for $N>1$ and some d? Can one find some system that is contextual for $N>1$, but is non-contextual for $N=1$? What are the consequences of a relaxation of the particle number conservation (for example for coherent states of photons)? In addition, one can study a quantum to classical transition by asking: does the system become non-contextual in the limit $N\rightarrow \infty$?

%%%%%%%%%%%%%%%%%%%%%%%%%%%%%%%%%%%%%%%%%%%%%%%%%

{\it Acknowledgements.} This work is supported by the National Science Centre in Poland through the NCN Grant No. 2014/14/E/ST2/00585.

%%%%%%%%%%%%%%%%%%%%%%%%%%%%%%%%%%%%%%%%%%%%%%%%%

\end{document}